\documentclass{emulateapj}
\slugcomment{Submitted to ApJ: July 19, 2019}

\shorttitle{Shocks in inhomogeneous plasmas}   %not more than 44 characters  ( confirmed in 18. 07. 2019.  )
\shortauthors{Tomita,  Ohira, and Yamazaki}
\begin{document}

\title{Weibel-Mediated Shocks Propagating into Inhomogeneous Electron-Positron Plasmas}

\author{Sara Tomita\altaffilmark{1,2}}
\email{tomisara@phys.aoyama.ac.jp}

\author{Yutaka Ohira\altaffilmark{2}}

\author{Ryo Yamazaki\altaffilmark{1}}

\altaffiltext{1}{Department of Physics and Mathematics, Aoyama Gakuin University, 
5-10-1 Fuchinobe, Sagamihara 252-5258, Japan}
\altaffiltext{2}{Department of Earth and Planetary Science, The University of Tokyo, 
7-3-1 Hongo, Bunkyo-ku, Tokyo 113-0033, Japan}

%% Note that the \and command from previous versions of AASTeX is now
%% depreciated in this version as it is no longer necessary. AASTeX 
%% automatically takes care of all commas and "and"s between authors names.

%% AASTeX 6.2 has the new \collaboration and \nocollaboration commands to
%% provide the collaboration status of a group of authors. These commands 
%% can be used either before or after the list of corresponding authors. The
%% argument for \collaboration is the collaboration identifier. Authors are
%% encouraged to surround collaboration identifiers with ()s. The 
%% \nocollaboration command takes no argument and exists to indicate that
%% the nearby authors are not part of surrounding collaborations.

%% Mark off the abstract in the ``abstract'' environment. 
\begin{abstract}

%The Weibel instability is kinetic plasma instability which generates the magnetic fields in anisotropic temperature plasmas, and it is expected to occur in various astrophysical high energy phenomena.
The external forward shock emitting the gamma-ray burst (GRB) afterglow is collisionless, and it is mediated by the Weibel instability which generates the magnetic field.
The GRB afterglow shows that the magnetic field in the large downstream region is much stronger than the shock-compressed pre-shock field.
However, particle-in-cell (PIC) simulations of relativistic shocks propagating into homogeneous media show that the Weibel generated field decays near the shock front.
Some GRB observations and theoretical studies suggest that the preshock medium is inhomogeneous.
We perform the PIC simulation of a relativistic shock propagating into inhomogeneous plasma. 
It is found that the post-shock magnetic field decays slowly compared with the homogeneous case. 
Sound waves and entropy waves are also generated by the shock-wave interaction, and temperature anisotropy is produced by the sound wave in the downstream region.
The free energy of the temperature anisotropy is large enough to explain the observed field strength.
Our results show that the upstream density fluctuation has a significant effect in the downstream region of collisionless shocks even if the wavelength of the upstream inhomogeneity is much larger than the kinetic scale.
% peak time of optical afterglows, 1 day.
%We expect that the second-order Fermi acceleration in the post-shock region for clumpy media. 

\end{abstract}

%% Keywords should appear after the \end{abstract} command. 
%% See the online documentation for the full list of available subject
%% keywords and the rules for their use.
%\keywords{gamma-ray burst: general \UTF{2013} instabilities \UTF{2013} magnetic fields \UTF{2013} plasmas \UTF{2013} shock waves }
\keywords{gamma-ray burst: general --- instabilities --- magnetic fields --- plasmas --- shock waves }

%% From the front matter, we move on to the body of the paper.
%% Sections are demarcated by \section and \subsection, respectively.
%% Observe the use of the LaTeX \label
%% command after the \subsection to give a symbolic KEY to the
%% subsection for cross-referencing in a \ref command.
%% You can use LaTeX's \ref and \label commands to keep track of
%% cross-references to sections, equations, tables, and figures.
%% That way, if you change the order of any elements, LaTeX will
%% automatically renumber them.
%%
%% We recommend that authors also use the natbib \citep
%% and \citet commands to identify citations.  The citations are
%% tied to the reference list via symbolic KEYs. The KEY corresponds
%% to the KEY in the \bibitem in the reference list below. 

\section{Introduction} \label{sec:intro}

The energy dissipation in relativistic shocks is important in the study of the production of relativistic
particles (cosmic rays) and the photon emissions  in high-energy astronomical transients. 
Such phenomena occur in dilute environments,  so that the relativistic shocks arise through the kinetic processes in collisionless plasmas. 
The Weibel instability, which occurs in the collisonless plasmas with temperature anisotropy \citep{weibel59}, 
has been actively investigated in this field.
It generates magnetic fields and would be important in the particle acceleration in the relativistic collisionless shocks 
\citep{medvedev99,spitkovsky08b,sironi13,matsumoto17}. 
The Weibel instability also works for the excitation of non-relativistic shocks and especially is crucial role in the 
particle acceleration \citep{kato10,matsumoto15}. 
Recently, the generation of the Weibel mediated shock has been also studied using high-power lasers
\citep{fox13,huntington15,ross17}.

Observations of gamma-ray burst (GRB) afterglows have shown that they arise from the synchrotron radiation from energetic electrons in post-shock regions \citep{sari98}. 
Various theoretical modelings have suggested that  the fraction of  the magnetic-field energy density converted from kinetic energy density, $\epsilon_{\rm B}$, 
widely ranges from $\sim 10^{-8}$ to $\sim10^{-2}$ with the mean of about $10^{-5}$, with an assumption that the density of $1\ {\rm cm}^{-3}$ \citep{santana14}.
The lower limit is still larger than 
the shock-compressed value of interstellar magnetic field ($\sim10^{-9}$), so that the field amplification is necessary.
 Moreover, since afterglows continue to be bright over a period of days to years, the field 
 should remain strong in flowing downstream \citep{gruzinov99, piran05, katz07}.
In principle, the value of $\epsilon_{\rm B}$ may be constrained from the observed ratio of the peaks in the photon energy spectrum of the synchrotron and inverse Compton emissions.
However, it  still has large uncertainty because  of the small number of events detected in GeV-TeV energy bands.
 Therefore, it is important  to theoretically estimate the value of $\epsilon_{\rm B}$ through the micro-processes at the shock  
for understanding of the radiation mechanism of the GRB afterglows.

So far, the nonlinear evolution of the magnetic field generated by the Weibel instability has been investigated for shocks propagating 
into homogeneous plasmas
\citep{silva03, kato05, chang08, keshet09, medvedev11, lemoine15, takamoto18, ruyer18}. 
However, it is currently unknown whether the Weibel generated field can survive long enough to explain
the GRB afterglow emission.
According to the two dimensional pair plasma shock simulation in \citet{chang08}, the generated field rapidly decays, and the decay rate is  $\epsilon_{\rm B} \approx (t\omega_{\rm pe})^{-1}$, where $\omega_{\rm pe}$ is the plasma frequency of electrons measured in the downstream rest frame.
This means that the Weibel field can not persist during the GRB afterglow 
emission lasting for $t\omega_{\rm pp}\sim10^{7}$, where $\omega_{\rm pp}$ is the plasma frequency of protons measured in the downstream rest frame \citep{sironi07}.
In the long-term simulation ($t\omega_{\rm pe}\sim10^4$) performed by \citet{keshet09}, 
the field is moderate in decaying
because a wavelength of the magnetic field becomes large 
owing to the high-energy particles accelerated at the shock. 
The decay rate also depends on the time evolution of the spectrum of the field fluctuation \citep{chang08,lemoine15}. 
\citet{chang08} gave the analytic form of the decay rate as, $\epsilon_{\rm B} \propto  (t\omega_{\rm pe})^{-2(p+1)/3}$, where {\it p} is the long-wavelength spectral index, and $p\simeq1/2$ at late time ($t\omega_{\rm pe}\sim10^3$). 

Previous local PIC simulations with periodic boundary conditions
have shown that the current filaments more easily merge in the three-dimensional case than in the two-dimensional case. 
The wavelength becomes longer and the decay rate decreases in both cases of 
the electron-positron \citep{silva03} and electron-ion plasmas \citep{takamoto18}. 
\citet{medvedev11} suggested that the decay exponent is $-2$ in the pair plasma, 
although the simulation time ($t\omega_{\rm pe}\simeq 50$) is much shorter than the timescale of GRB afterglows.
Recent three-dimensional simulation with larger simulation box in the flow direction has shown 
that the current filament in electron-ion plasma is broken by the drift-kink mode before being large scale \citep{ruyer18}.  
These simulations were, however, performed with the periodic boundary condition along the flow direction,
which is different from the non-periodic shock structure, and no effect  of particle acceleration is considered.

Meanwhile, the observed large dispersion of $\epsilon_{\rm B}$ appears to be a key ingredient in understanding the origin of the amplified magnetic field.
If the field strength and plasma temperature in upstream region are similar to those in Galactic ISM, 
then both Alfv\'{e}n  and sonic Mach numbers are much larger than unity 
at early afterglows, where we assume the Lorentz factor of the flow $\sim100$. 
Then, the downstream shock structure does not depend on these quantities,
that is,  they are not important parameters in calculating the GRB afterglow emission \citep{sari98}.
More important is the upstream number density to determine the downstream magnetic field.
\citet{santana14} showed that  the amplification of the magnetic field has the weak dependence of the upstream density; ${\rm AF}\propto n_{\rm 0}^{0.21}$, where AF is an amplification factor characterizing how mach the seed magnetic field was amplified beyond the shock compression.
Hence, one can see that the value of $\epsilon_{\rm B}$ is not determined by the upstream density alone, and there should be another piece controlling the downstream state.
 
%一方で、上流の数密度は、下流磁場を決める重要なパラメターの一つである。
%観測されている\epsilon_Bの分布は6桁にもわたり、それは上流の数密度の分布に対応するかもしれない。
%しかし、上流の数密度の分布にそれほどの広がりは、これまでに確認されたたことはない。
%santana14は、磁場増幅には、上流密度の依存性は弱いと言っている(AF\propton^{0.21}、AFはamplification factorで、種磁場が、衝撃波圧縮以外に、どれほど増幅されているかを示す)。
%したがって、\epsilon_Bの値は、上流密度だけでなく、他に、下流の状態を決める重要なパラメターがあるはずである。

% In addition to the upstream density, the wavelength of the upstream density fluctuation 
% is unknown scales, so that the unshocked density fluctuation is expected to be one of the parameter introducing a new scale to the shocked magnetic fields. 

The upstream medium of the shock wave is generally inhomogeneous.  
For example, a progenitor of the long GRB is thought to be a massive star. 
Then, the  environment around them has some density fluctuations produced by the stellar wind  or some inhomogeneity of interstellar medium (ISM). 
The characteristic length scale of the density fluctuations is uncertain;
for example, it is on the order of 100~pc for ISM turbulence \citep{armstrong95}, 
or smaller for circumstellar matter (CSM) \citep{chugai94,smith09,yalinewich19}. 
There are some events suggesting the upstream density fluctuations \citep{wang00, lazatti02, dai02, heyl03, schaefer03, nakar03}.

Then, the large dispersion of $\epsilon_{\rm B}$ may come from the inhomogeneity of the surrounding environment, that is,
the wavelength of the upstream fluctuation potentially becomes an important parameter in describing the downstream magnetic field.
In fluid scale, the field amplification process depends on the environment around a progenitor star.
Magnetohydrodynamics (MHD) simulations have shown that the magnetic field is amplified by turbulent dynamo when the shock is propagating into the inhomogeneous media \citep{sironi07, inoue11, mizuno11}. 
In this case, the decay timescale of the field is the eddy turn over time.

However, when the turbulent kinetic energy is dissipated, particle diffusion or kinetic plasma instability is excited in collisionless plasma. 
In addition, diffusive motion of high-energy particles may potentially impact on the shock structure.
With fluid approximation, 
a linear analysis of an interaction between a shock and an incident entropy wave shows that entropy and  subsonic modes
are excited by the mode conversion in front of and behind the shock \citep{mckenzie68, lemoine16}. 
\citet{shu89} and \citet{arshed13}
studied a shock tube problem in the case of the interaction between the shock and some waves. 
They claimed that an upstream incident entropy wave is compressed at the shock front, 
being advected stably  in the downstream region. 
However, it is not clear whether the incident waves passing through the shock front can persist in kinetic scale when
the particle diffusion is non-negligible. 
In this case, the density fluctuations may rapidly decay before the magnetic field is amplified by the turbulence in large scale.
 Moreover, the energy dissipation by the particle diffusion is expected to lead to the plasma heating and particle acceleration. 
The kinetic plasma instability may convert the turbulent kinetic energy to the magnetic energy. 
In this paper, using first principles approach, we investigate possible effects of the upstream density fluctuation on the evolution of downstream magnetic field.

In our previous study \citep{tomita16},
two-dimensional local PIC simulation was performed to see  the nonlinear evolution of the Weibel instability in electron-positron plasma with an anisotropic density structure, 
which mimics the downstream region of a relativistic shock propagating into inhomogeneous media. 
We showed that particles escape from an anisotropic high-density region, leading
a temperature anisotropy large enough for the Weibel instability to make the magnetic field fluctuation.
 
This paper is organized as follows.
In Section~\ref{sec:2.1}, we conduct two-dimensional PIC simulation of relativistic shocks propagating into the inhomogeneous media for the first time.  
Our simulation shows that the Weibel magnetic field has larger scale and higher magnitude than that for the case of the homogeneous density distribution. Such magnetic fields can persist in the far downstream region without significant decay.
We also confirmed there are the entropy and sound waves in the downstream region during the simulation time.
The temperature anisotropy is generated by the diffusion of high-energy particles, being maintained in the far downstream region. 
The field amplification by the Weibel instability is however, less significant in the downstream region (Section~\ref{sec:2.2}).  
It is finally claimed in Section~\ref{sec:3} that the measured temperature anisotropy is large enough to explain the observed GRB afterglows. 
We also discuss the wavelength of the density fluctuation around the progenitor of GRBs (Section~\ref{sec:3}).

\section{simulation setting} \label{sec:2.1}

We use the two-dimensional electromagnetic PIC code, pCANS. 
In order to suppress the numerical Cherenkov instability \citep{godfrey74}, Maxwell's equations are solved by the implicit method with the Courant-Friedrichs-Lewy number of 1.0 \citep{ikeya15}. 
We set a two-dimensional simulation box in the $x$-$y$ plane. Periodic boundary condition is imposed in the $y$-direction. 
The simulation box size is $L_x = 1.2 \times 10^4 c/\omega_{\rm pe}$ and $L_y = 86.4 c/\omega_{\rm pe}$, where 
\begin{equation}
\frac{c}{\omega_{\rm pe}}=\sqrt{\frac{\Gamma m_{\rm e}c^2}{4\pi n_{\rm 0} e^2}} ,
\end{equation}
is the plasma skin-depth of the upstream electron-positron plasma with a bulk Lorentz factor $\Gamma$ and upstream mean number density $n_{\rm 0}$. Constants $c, e, m_{\rm e}$ are speed of light, particle charge, particle mass, respectively.
The cell size and time step are $\Delta x=\Delta y=0.1c/\omega_{\rm pe}$ and $\Delta t=0.1\omega_{\rm pe}^{-1}$, respectively. 

Unmagnetized plasma with the bulk Lorentz factor of $\Gamma=10$ and the thermal velocity of $v_{\rm th}=0.1c$ is injected from the left boundary ($x=0$) and reflected at the right boundary ($x=1.2\times 10^4 c/\omega_{\rm pe}$). 
The reflected and incoming particles interact, forming a shock which propagates toward the injection wall. 
The simulation frame is the downstream rest frame, that is, the downstream mean velocity is zero. 
We use 40 particles per cell per species. 
 We investigated a dependence of a numerical heating on the number of particles by performing simulations for 10, 20, and 40 particles per cell per species. 
We confirmed that the more particles, the more suppressed are the numerical heating. 
We show simulation results for 40 particles per cell per species in this paper.

To investigate a collisionless shock propagating to an inhomogeneous plasma, the spatial distribution of the electron-positron plasma in the upstream region is set to be
\begin{equation}
n(x,y) = n_0[1+0.5 \sin(2\pi x/\lambda_x)].
\end{equation}
Namely, the density fluctuation is given by a monochromatic wave with $\delta n/n_0 = 0.5$. 
For simplicity, the upstream temperature is set to be constant, which makes the pressure fluctuation but it is negligible because the shock Mach number is very large. 
We performed two runs for $\lambda_x = 4\times10^2 c/\omega_{\rm pe}$ and $1.2 \times10^3 c/\omega_{\rm pe}$ 
in order to study a wavelength dependence.
We confirmed that our results did not significantly depend on $\lambda_x$. 
In the following, we show simulation results for $\lambda_x = 1.2\times10^3 c/\omega_{\rm pe}$ which is much larger than the shock thickness of the Weibel mediated shock, $\sim 50 c/ \omega_{\rm pe}$.

%fig1
\begin{figure*}%[htbp]
%\begin{center}
%\includegraphics[width=18.0cm]{shockstructure2.eps} %fig1all-nou.eps}     \\
\plotone{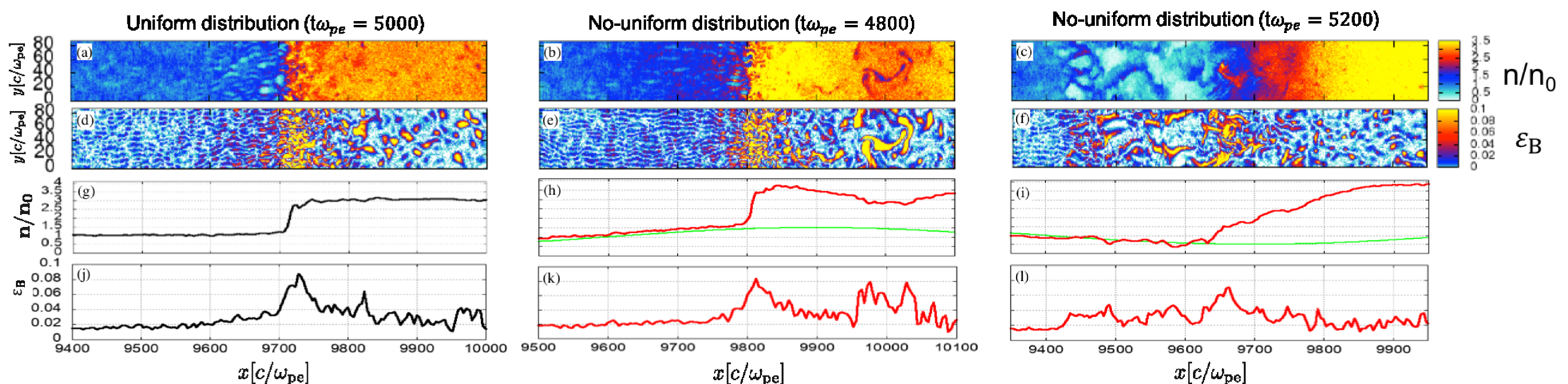}
%\vspace*{+0.5cm}
\caption{
Structures of shocks for the homogeneous density distribution at $t\omega_{\rm pe}=5000$ (left panels), and inhomogeneous case at  $t\omega_{\rm pe}=4800$ (middle panels) and $5200$ (right panels). 
At $t\omega_{\rm pe}=4800$ and 5200, the shock front is passing through the highest- and lowest-density regions, respectively. 
Panels (a)-(c) show the plasma density normalized by the upstream mean plasma density, $n/n_0$. 
Panels (d)-(f) show the magnetic energy density normalized by the upstream mean kinetic energy density, $\epsilon_{\rm B}=B^2/32\pi\Gamma n_0 m_{\rm e} c^2$. 
Panels (g)-(i) and panels (j)-(l) show transversely averaged $n/n_0$ and $\epsilon_{\rm B}$, respectively.
The green curves in panels (h) and (i) show the initial density profile for the inhomogeneous case.
 \label{fig1}}
%\end{center}
\end{figure*}

%%%%%%%%%%%%%%%%%%%%%%%%%%%%%%%%%%%%%%%%%%%%%%%%%%%%%%%
%%%%%%%%%%%%%%%%%%%%%%%%%%%%%%%%%%%%%%%%%%%%%%%%%%%%%%%

\section{result} \label{sec:2.2}
\subsection{shock structure}

First, we compare the shock structure for the case of homogeneous and inhomogeneous density distributions.  
The left column in Figure~\ref{fig1} shows the shock structures for the homogenous case at $t\omega_{\rm pe}=5000$. 
From top to bottom, the four panels show the two-dimensional structures of normalized electron-positron density $n/n_0$ and the energy fraction of the magnetic field $\epsilon_{\rm B}=\rm{B^2}/( 32\pi \Gamma n_0 m_{\rm e} c^2)$, and one-dimensional profiles averaged over the $y$-direction of $n/n_0$ and $\epsilon_{\rm B}$, respectively.
The shock front is located at $x\approx 9700~c/\omega_{\rm pe}$ and moving leftward with the velocity of $v_{\rm sh}\approx0.37~c$. 
The left side of the shock front is the upstream region where magnetic filaments are generated by the Weibel instability. 
The transverse width of the Weibel filaments is about $8~c/\omega_{\rm pe}$ in the upstream region. 
The magnetic field generated by the Weibel instability decays from the shock front downstream. 
These features are consistent with early studies for relativistic Weibel mediated shocks \citep{spitkovsky08b,keshet09}. 
Compared with our previous periodic simulation \citep{tomita16} in which the Weibel instability is excited by counter-streaming plasmas, the Weibel-mediated field is maintained for a long time in the shock simulation. 
This is because the magnetic field with a longer wavelength can be excited in the shock simulation. 
According to \citet{keshet09}, the large-scale magnetic field is generated by high-energy particles accelerated by the shock.

%%%%%%%%%%%%%%%%%%%%%%%%%%%%%%%%%%%%%%%%%
%%%%%%%%%%%%%%%%%%%%%%%%%%%%%%%%%%%%%%%%%

The middle and right panels in Figure~\ref{fig1} display the time sequence of the shock structures for the inhomogeneous case at $t\omega_{\rm pe} = $4800 and 5200, respectively. 
At $t \omega_{\rm pe}= 4800$ (middle panels in Figure~\ref{fig1}), the shock front is passing through the highest-density region. 
Although the upstream structures are quite similar to those for the homogeneous case, one can see significant differences in the downstream region. 
At $x \approx 10000\ c/\omega_{\rm pe}$, there are large-scale filaments and the magnetic-field strength is comparable to that in the shock front. They are generated as the shock front passes through the lower-density region. 
The right column in Figure~\ref{fig1} shows the shock structures at $t\omega_{\rm pe}=5200$ when the shock front is passing through the lowest-density region. 
The size of the upstream filaments is significantly larger than that for the homogeneous case. 
In addition, the magnetic field is widely generated in the broad upstream region (see panels (f) and (l)). 
The Weibel instability is driven by the cold upstream plasma and the hot plasma leaking from the downstream region. 
Since the density ratio of the hot leaking plasma to the cold upstream plasma becomes larger in the low-density region, 
the large momentum dispersion of the hot leaking plasma makes the transverse width of the filament wider \citep{ruyer17}. 
Since the length scale of the Weibel magnetic field is larger when the shock propagates in lower-density regions, 
it can be advected far downstream without dissipation. 
This is the reason why there is large-scale magnetic field in the downstream region at $t\omega_{\rm pe}=4800$ (panels (e) and (k)).

\begin{figure}[htbp]
%\begin{center}
%\includegraphics[keepaspectratio, width=13.0cm]{b1d.eps} %width=10.0cm]{fig1-1.eps}     \\
\plotone{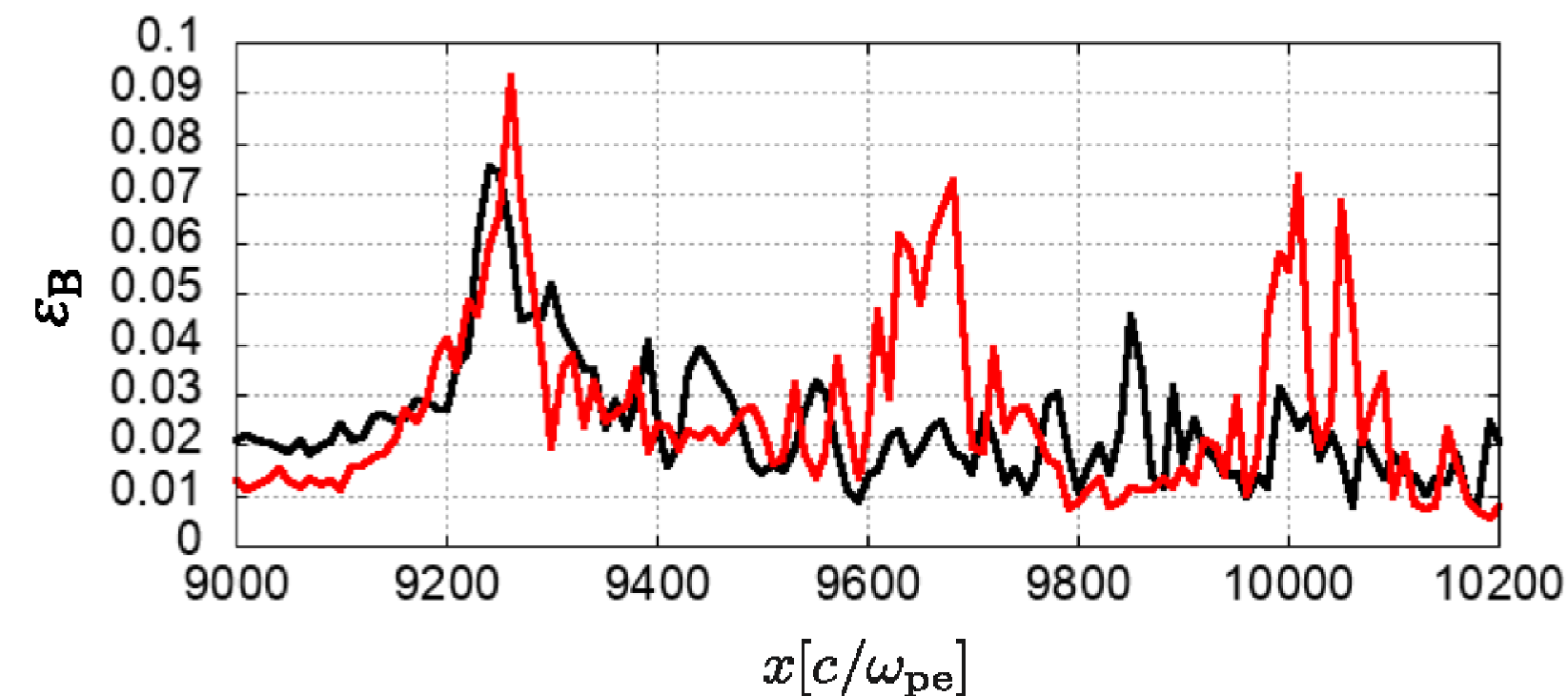}
%\vspace*{+1.2cm}
\caption{Spatial profiles of the transversely averaged $\epsilon_{\rm B}=B^2/32\pi\Gamma n_0 m_{\rm e} c^2$ in the larger downstream region at $t\omega_{\rm pe}=6100$ for the homogeneous (black) and inhomogeneous (red) density distributions, respectively. The shock front is located at $x \approx 9200~c/\omega_{\rm pe}$. 
\label{fig2}}
%\end{center}
\end{figure}

%%%%%%%%%%%%%%%%%%%%%%%%%%%%%%%%%
%%% magnetic field in the large downstream region %%%
In Figure~\ref{fig2}, we compare the spatial profile of the transversely averaged $\epsilon_{\rm B}$ in larger downstream region at $t\omega_{\rm pe}=6100$ for the homogeneous (a) and inhomogeneous (b) density distributions, respectively. The shock front is located at $x \approx 9200~c/\omega_{\rm pe}$ and passing through the lower-density region for the inhomogeneous case. 
For the homogeneous case, the magnetic field generated in the shock transition region simply decays downstream.
For the inhomogeneous case, we can see two peaks of $\epsilon_{\rm B}$ downstream, which decay much slower than that for the homogeneous case. 
The number of the peaks is the same as that of the waves which transmitted through the shock. 
Therefore, the inhomogeneity of the upstream density has an important role on the magnetic field generation far downstream. 

%fig2

%%%%%%%%%%%%%%%%%%%%%%%%%%%%%%%%%
%fig3
\begin{figure}[htbp]
%\begin{center}
%\includegraphics[width=9.0cm]{fig_denvx.eps} 
\plotone{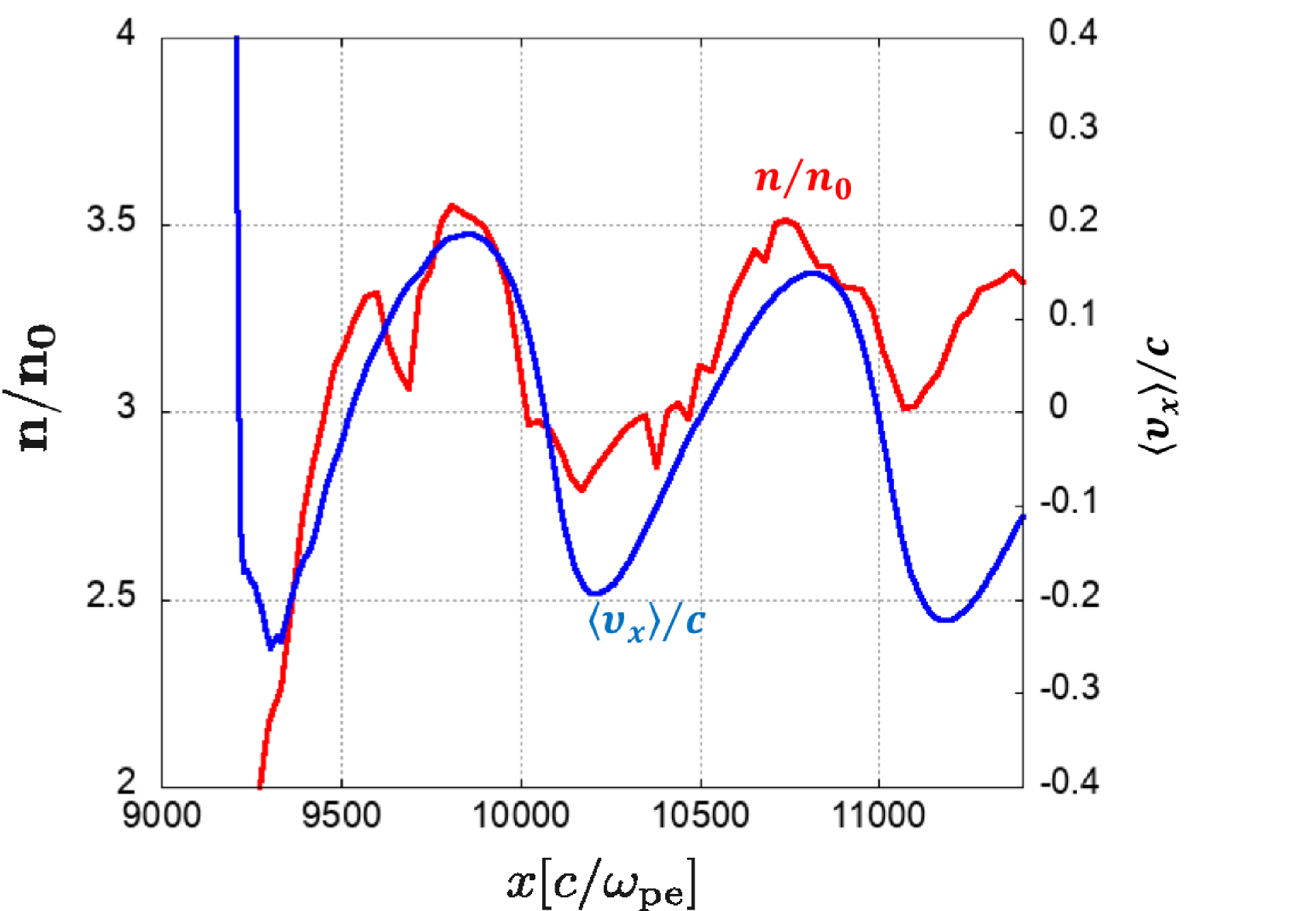} 
%\vspace*{+1.5cm}
\caption{Spatial profiles of the $x$ component of the average velocity, $\langle v_x\rangle / c$ (blue), and density, $n/n_0$ (red), measured in the downstream rest frame at $t\omega_{\rm pe}=6100$ in the downstream region for the inhomogeneous upstream case. The shock front is located at $x \approx 9200~c/\omega_{\rm pe}$. 
\label{fig3} }
%\end{center}
\end{figure}

Next, we concentrate on the large-scale downstream structures of density and velocity for the inhomogeneous case. 
The blue and red curves in Figure~\ref{fig3} show the one-dimensional structures of $x$ component of the average velocity, $\langle v_x\rangle / c$, and density, $n/n_0$, at $t\omega_{\rm pe}=6100$ when the shock front is located at $x \approx 9200~c/\omega_{\rm pe}$. 
As seen in the blue curve, the average velocity oscillates with an amplitude of $\sim 0.2 c$.
The velocity profile (the blue curve) almost correlates with the large-scale density fluctuation. 
Therefore, the sinusoidal variation with the wavelength $\lambda_{\rm down} \approx 1100~c/\omega_{\rm pe}$ can be interpreted as a sound wave propagating downstream. 
The propagation velocity in the downstream rest frame is about ${\rm 0.62c}$ that is consistent with the sound velocity of  relativistic plasmas, $c/\sqrt{3}$. 

There is another intermediate-scale density fluctuation with a wavelength of about $400~c/\omega_{\rm pe}$ at $x \approx$ 9500, 10000, 10500, 10900 $c/\omega_{\rm pe}$ in Figure~\ref{fig3}, which is not associated with the velocity displacement. 
They are non-propagating.
Therefore, this density fluctuation can be interpreted as an entropy wave. 
The observed wavelengths of the entropy and sound waves are consistent with results from hydrodynamic analysis of interaction between a shock wave and a monochromatic density wave. 
This should never be taken for granted because we consider a collisionless system at finite temperature. 
If no particles change their momentum, density structures have to disappear in a sound crossing time. 
However, the simulation indicates that entropy and sound waves do not decay in the downstream region.
The Weibel magnetic field changes the particle's momentum and plays an important role in the momentum exchange of particles.
The Larmor radius of thermal particles $r_{\rm g}$ is comparable to or slightly smaller than the coherence length of the Weibel magnetic field fluctuation, so that the diffusion coefficient of the thermal particles can be approximately given by
\begin{equation}
\kappa \sim \frac{1}{3}r_{\rm g}c=\frac{1}{3\sqrt{8}}  \frac{c^2}{\omega_{\rm pe}} \epsilon_{\rm B}^{-1/2} ~~,     \nonumber 
\end{equation}
%
%where we use the approximation which the dependence of the Lorentz factor of thermal particles is equal to the bulk Lorentz factor.
Then, the decay time scales due to diffusion of the entropy and sound waves in the downstream region are given by 
\begin{eqnarray}
t_{\rm decay}& = &\frac{\lambda_{\rm down}^2}{2\kappa} \nonumber \\   
&=&\left\{ \begin{array}{ll}
9.9\times 10^5\ \omega_{\rm pe}^{-1}  \left(\frac{\epsilon_{\rm B}}{10^{-2}}\right)^{1/2} \left( \frac{\lambda_{\rm down}}{1.1\times  10^3 c/\omega_{\rm pe}}\right) ^2  \\
~~~~~~~~~~~~~~~~~~~~~~~~~~~({\rm  for\ sound\ wave}) \\
 1.3\times 10^5\ \omega_{\rm pe}^{-1}  \left(\frac{\epsilon_{\rm B}}{10^{-2}}\right)^{1/2}\left( \frac{\lambda_{\rm down}}{4.0 \times 10^2 c/\omega_{\rm pe}}\right) ^2 \\
~~~~~~~~~~~~~~~~~~~~~~~~~({\rm for\ entropy\ wave}). 
 \end{array} \right. 
\end{eqnarray}
The decay time scales of the sound and entropy waves are much longer than the simulation time, $t\omega_{\rm pe}= 6100$ and sound crossing time, $t\omega_{\rm pe}=1935$. 
Therefore, the both waves can propagate in the downstream region during the simulation time. 
%

%fig4
%
\begin{figure}%[htbp]
%\begin{center}
%\includegraphics[width=12.0cm]{fig-tavx.eps}     
\plotone{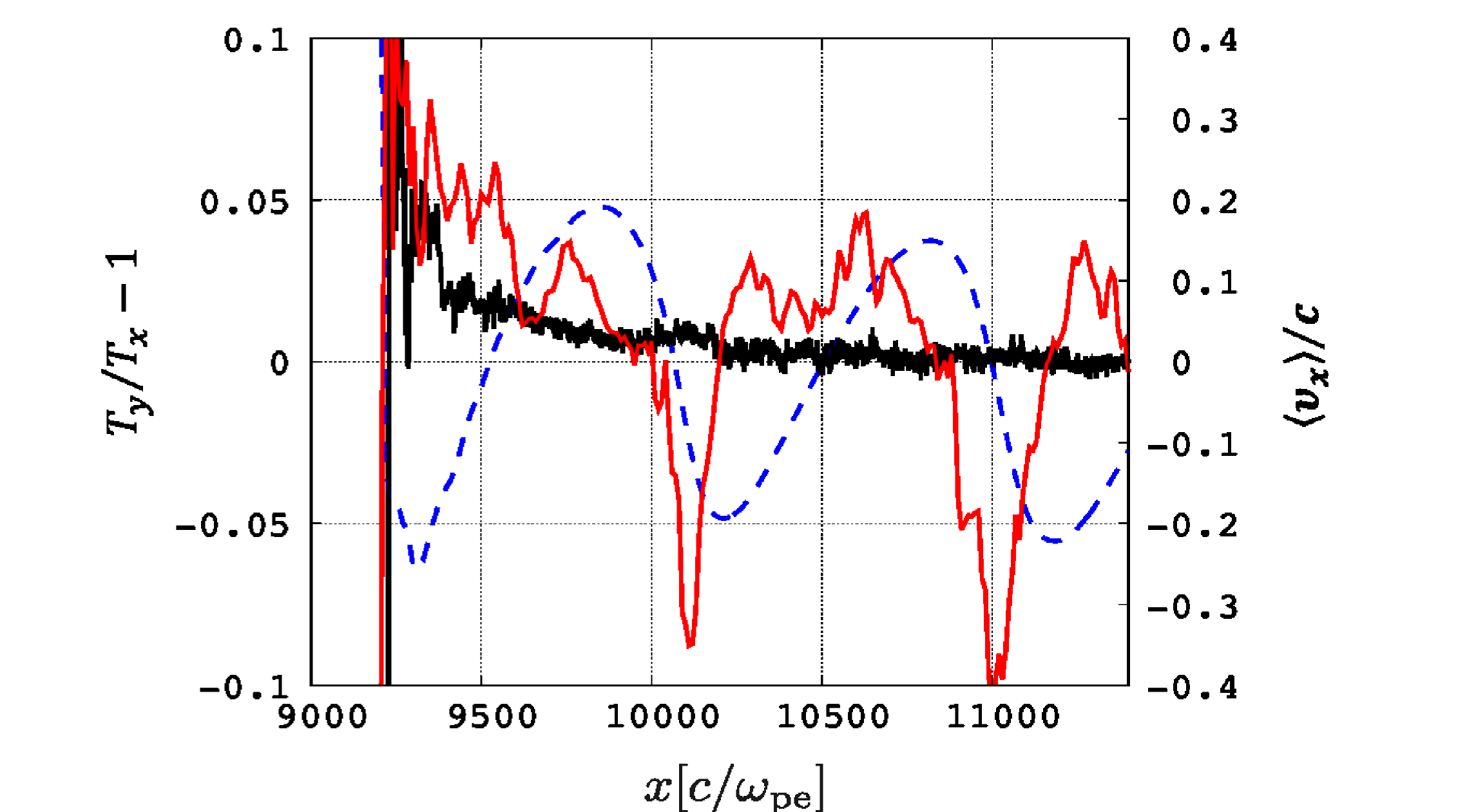}   
%\vspace*{+1.3cm}
\caption{ Spatial profiles of the temperature anisotropy at $t\omega_{\rm pe}=6100$ in the downstream region for the case of homogeneous (black) and inhomogeneous (red) upstream density distribution, respectively. The blue dashed curve shows the $x$ component of the average velocity for the inhomogeneous case, $\langle v_x\rangle / c$  measured in the downstream rest frame. The shock front is located at $x \approx 9200~c/\omega_{\rm pe}$. 
\label{fig4}}
%\end{center}
\end{figure}
%

%%%%%%%%%%%%%%%%%%%%%%%%%%%%%%%%%%%%%%%%%%%%%%%%%%%%%%%
%%%%%%%%%%%%%%%%%%%%%%%%%%%%%%%%%%%%%%%%%%%%%%%%%%%%%%%
\subsection{temperature anisotropy}

Figure~\ref{fig4} shows the spatial profile of temperature anisotropy averaged over the $y$-direction in the downstream region for the homogeneous (black solid) and inhomogeneous (red solid) cases at $t\omega_{\rm pe}=6100$. 
The blue dashed curve shows the $x$ component of the average velocity, $\langle v_x\rangle / c$.
The temperature anisotropy is defined by $T_y/T_x -1$, where according to \citet{yoon07} the temperatures $T_x$ and $T_y$ are given by 
\begin{eqnarray}
T_x &=& \int {\rm d}^3 p \frac{p_x^2}{\gamma m_e} f(p_x, p_y, p_z ),  \\
T_y &=& \int {\rm d}^3 p \frac{p_y^2}{\gamma m_e} f(p_x, p_y, p_z ). 
\end{eqnarray}
Here, $\gamma, f, p_x, p_y,$ and $p_z$ are the particle Lorentz factor, the distribution function and momentum in the center-of-mass rest frame which is calculated by all particles in region of width $\sim 20 c/\omega_{\rm pe}$. 
The temperature anisotropy is almost zero in the downstream region for the homogeneous case, 
while for the inhomogeneous upstream case it oscillates synchronously with the sound wave with values ranging from $-0.1$ to 0.05. 
%it has values from $-0.1$ to $0.05$ and oscillates synchronously with the sound wave for the inhomogeneous case. 
It has been confirmed that the amplitude of the temperature anisotropy does not change even when the wavelength of upstream fluctuations becomes three times smaller than that in this simulation. 
As seen in Figure~\ref{fig4}, the absolute value of the temperature anisotropy becomes maximum at $x \approx 10100\ c/\omega_{\rm pe}$ and $ \approx 11000\ c/\omega_{\rm pe}$ where the divergence of the mean velocity is negative. 
The origin of the temperature anisotropy will be discussed later.

The temperature anisotropy can generate magnetic field by the Weibel instability in the downstream region. 
However, the observed temperature anisotropy oscillates downstream. 
We need to compare the growth time scale of the magnetic field and oscillation period of the temperature anisotropy in order to investigate whether the magnetic field is generated by the Weibel instability or not. 
The maximum growth rate of the relativistic Weibel instability is given by,
\begin{eqnarray}
\gamma_{\rm max} &=& \frac{2}{3\sqrt{3}} \frac{T_x}{m_{\rm e}c^2} \left( \frac{T_y}{T_x} -1 \right)^{3/2} \Gamma^{1/2} \omega_{\rm pe}~~, \label{eq7}%\\
%k_{\rm max}&=&\frac{1}{3}\sqrt{ \alpha_x \left(\frac{\alpha_x }{ \alpha_y} -1\right)} \frac{\omega_{\rm pe}}{c}~~,
\end{eqnarray}
where $T_y > T_x$ and $\gamma_{\rm max} \ll \omega_{\rm pe}$ are assumed \citep{yoon07}. 
For the temperature anisotropy of 0.05 observed in our simulation, the growth time scale is 
\begin{eqnarray}
t_{\rm grow} &=& \gamma_{\rm max}^{-1} \nonumber \\
         &\approx&   20\ \omega_{\rm pe}^{-1} \left(\frac{T_x/m_{\rm e}c^2}{3.8}\right)^{-1} \nonumber \\
       &~~&\times   \left(\frac{T_y/T_x-1}{0.05}\right)^{-3/2} \left(\frac{\Gamma}{10}\right)^{-1/2} \nonumber .
\end{eqnarray}
The oscillation period of the temperature anisotropy is that of the sound wave, $\lambda_{\rm down}/c_{\rm s}\approx1.7\times10^3 \omega_{\rm pe}^{-1}$. 
Since $t_{\rm grow} < \lambda_{\rm down}/c_{\rm s}$, the Weibel-unstable mode should grow in the downstream anisotropic temperature plasma. 
However, we could not identify the magnetic field generated by the Weibel instability in the downstream region. 
%式(7)は、背景磁場がない時の線形解析から得たワイベル不安定性の成長率のため、シミュレーションでは、成長率が衝撃波面で作られた下流磁場によって抑えられている可能性もある。
%また線形解析はMaxwell Juttner分布を仮定しているが、simulationの非等方性は、nonthermal particleが作っているので、合わないのかも。
%もし成長時間 $2\times 10^{2} \omega_{\rm pe}^{-1}$の間で、温度非等方性のエネルギーが全て磁場エネルギーに渡っても、この約数パーセントに保たれている下流磁場の中では、下流ワイベル不安定性の優位な成長は見られない。
This might be because equation (\ref{eq7}) is given by the linear analysis of the Weibel instability in unmagnetized plasma, so that the growth rate is suppressed by the downstream magnetic field generated in the shock front. 
In addition, although the linear analysis of the Weibel instability also assumed the Maxwell-J$\ddot{u}$ttner distribution,
the observed temperature anisotropy is produced by the non-thermal particles.
This could also be the reason for the discrepancy in the growth rate.
Even if the free energy of the temperature anisotropy is fully converted to the magnetic-field energy during $t_{\rm grow}\approx 20 \omega_{\rm pe}^{-1}$,
we cannot see it significantly in the downstream field of the strength keeping a few percent.
%This is because the downstream field generated in the shock front already keeps the energy fraction of more than ten  percent. 
%Even if the free energy of the temperature anisotropy is fully converted to the magnetic-field energy, it cannot exceed the field strength advected from the shock front. 
If the magnetic field generated in the shock front sufficiently decays in the far downstream region, 
the observed temperature anisotropy could play an important role in the generation of the magnetic field in the far-downstream region. 
%
%

%%%%%%%%%%%%%%%%%%%%%%%%%%%%%%%%%%%%%%%%%

%%% the origin of the temperature anisotropy
%fig5
\begin{figure}%[htbp]
%\begin{center}
%\includegraphics[scale=0.45]{fig-cartoon.eps} %[width=7.5cm]{fig-cartoon.eps}
\plotone{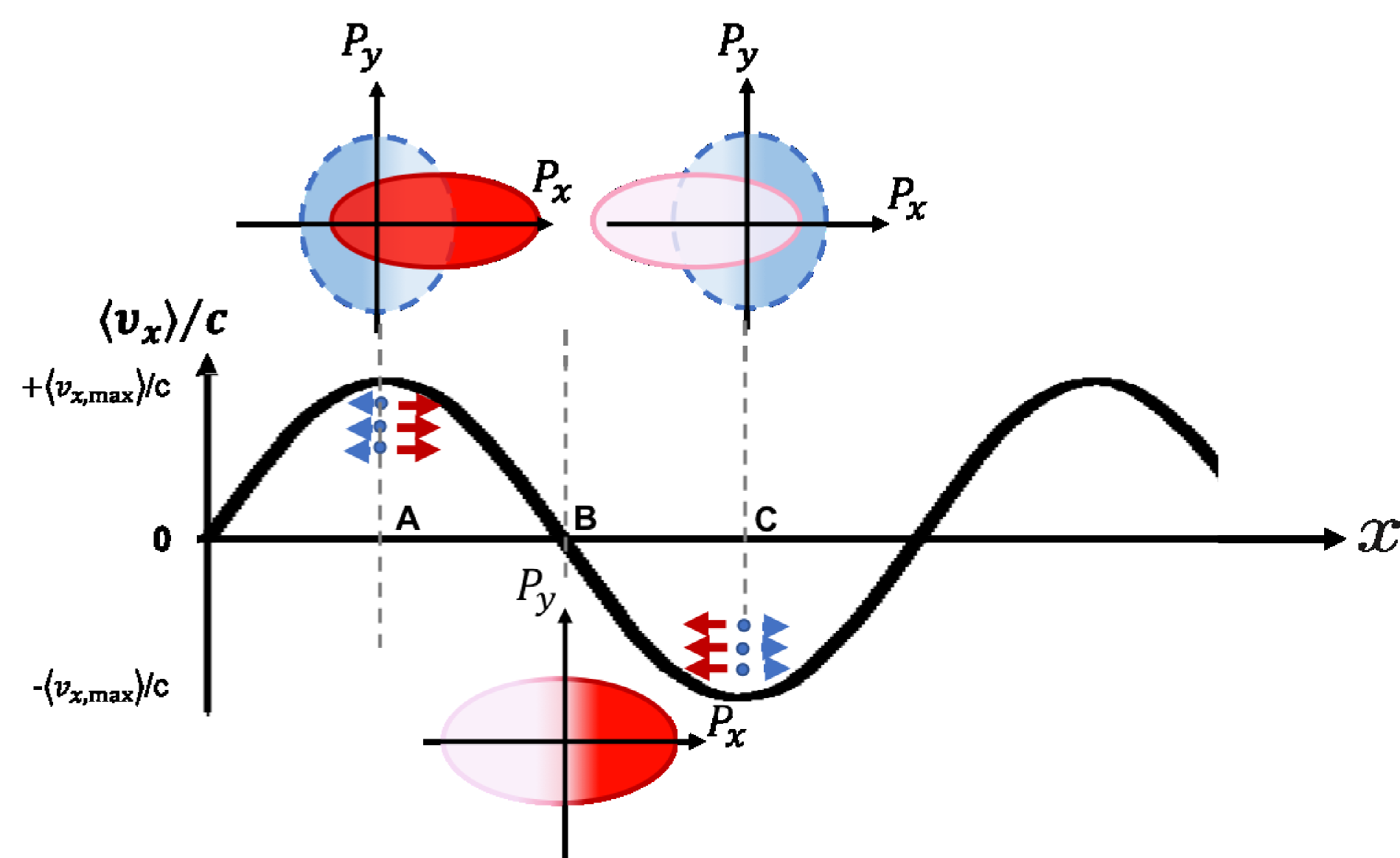}
%\vspace*{+1.0cm}
\caption{Generation mechanism of the temperature anisotropy in the downstream region. 
The blue dashed circles are isotropic momentum distribution in the rest frame of regions A and C. 
The red solid circles are the momentum distribution measured in the rest frame of region B (downstream rest frame). The black curve is the spatial profile of $x$ component of the average velocity, $\langle v_x\rangle / c$.
\label{fig5} }
%\end{center}
\end{figure}
We explain the origin of the downstream temperature anisotropy. 
As mentioned above, the particle diffusion in the sound wave is negligible for thermal particles. 
On the other hand, for high-energy particles accelerated by the shock, their gyroradii become larger than the coherent length scale of the magnetic-field fluctuations, so that the diffusion of high-energy particles becomes important. 
Therefore, the mean motion of the high-energy particles deviates from the sound wave motion. 
Figure~\ref{fig5} shows the cartoon illustrating the momentum distribution in $p_x - p_y$ plane at the regions A, B, and C, where mean velocities are $\langle v_x\rangle > 0, = 0$, and $< 0$, respectively.
The blue dashed circles are isotropic momentum distribution in the rest frame of regions A and C. 
If they are measured in the rest frame of region B, the momentum distributions are represented as the red circles by the Lorentz transformation. 
The color contrast (dark or light) represents the density contrast (high or low). 
Since the sound wave motion concentrates on particles in the region B, high-energy particles in the regions A and C are more concentrated to the region B. 
As a result,  the momentum distribution in the region B has larger momentum dispersion along the $p_x$-direction, so that the temperature in the $x$-direction becomes larger than that in the $y$-direction. 
On the other hand, the regions A and C lose particles with large $p_x$, so that, the temperature in the $x$-direction becomes smaller than that in the $y$-direction. 
Therefore, the temperature anisotropy in the downstream region is generated by the sound wave motion and diffusion of high-energy particles. 

\section{Discussion} \label{sec:3}

In this paper, we have investigated effects of upstream inhomogeneity on unmagnetized relativistic collisionless shocks by means of PIC simulations. 
In our simulation, an upstream density sinusoidaly changes with a wavelength of $1.2\times 10^3\ c/\omega_{\rm pe}$ in the downstream rest frame, which is much larger than the thickness of the shock transition region, the characteristic length scale of the Weibel instability, and the kinetic scale of plasmas. 
Although it is still much smaller than the injection scale of the interstellar turbulence ($\sim 1-100\ {\rm pc}$), there are some mechanisms that generate the small-scale density fluctuations in a precursor region of collisionless shocks \citep{drury86,ohira13b,ohira14,ohira16}. In addition, stellar winds contain small-scale density variation compared with the ISM turbulence. These small-scale density fluctuations could play an important role in collisionless shocks. 

Our simulation shows that, compared with results for shocks into a uniform plasma, a larger-scale magnetic field is generated in the shock precursor region when the shock propagates into a low-density region in the inhomogeneous plasma. The density ratio of the hot leaking plasma to the upstream cold plasma becomes larger in the low-density region.  Then, a finite temperature effect on the Weibel instability becomes significant in the low-density region, so that the characteristic length of the Weibel instability is thought to be larger. Hence, the larger-scale magnetic field hardly decays after they were advected downstream.  
The long-lived large-scale field is anticipated in GRB afterglows and 
could make a particle acceleration more efficient although this effect was not observed in this simulation because of a short simulation time. 
If we consider a longer length scale of upstream density fluctuations, 
the coherence length and the decay time in the downstream region of generated magnetic fields would be longer.

We found that the upstream density variation generates sound waves in the downstream region. 
The downstream sound waves are responsible for the temperature anisotropy which can generate the magnetic field by the Weibel instability. 
Therefore, the energy of the downstream sound waves is expected to be converted to the downstream magnetic-field energy. 
Our simulations have shown that the ratio of the sound-wave energy density $4n_0\Gamma m_{\rm e}\langle v\rangle^2/2$ to the kinetic energy density is  
\begin{eqnarray}
\epsilon_{\rm sw} %&=& \frac{\frac{1}{2}4n_0\Gamma m_{\rm e}\langle v\rangle^2}{4n_0\Gamma m_{\rm e}c^2} \nonumber \\
                       &\sim& 0.02 \times \left( \frac{\langle v\rangle}{0.2c} \right)^2 ~~, \label{eq3.1}
\end{eqnarray}
where the downstream density is assumed to be $4n_0$ and $n_0$ is the upstream number density in the downstream rest frame. 
Furthermore, our simulation have shown that temperature anisotropy on the order of $10^{-2}$ is generated in the downstream region by the sound wave, which means that the ratio of the free energy of the temperature anisotropy to the downstream kinetic energy is $\epsilon_{\rm ani}\sim10^{-2}$. 
Since observations of GRB afterglows suggest the energy fraction of magnetic field with $\epsilon_{\rm B}\sim10^{-5}$ \citep{santana14} which is smaller than $\epsilon_{\rm sw}$ and $\epsilon_{\rm ani}$, the energies of the downstream sound wave and temperature anisotropy are sufficient to generate the magnetic field in the afterglow emission region. 
Therefore, for shocks in inhomogeneous plasmas, 
the magnetic field generation takes place not only at the shock precursor region but also far downstream of the shock.
%in addition to the large-scale long-lived magnetic field generated in the shock precursor region, small-scale magnetic fields could be generated by the Weibel instability in the far downstream region. 

In order for the upstream density fluctuations to affect the GRB afterglow emission, the wavelength, $\lambda^{\rm u}$ measured in the upstream frame, must be smaller than the deceleration radius of GRB jet, $R_{\rm dec}^{\rm u}$, that is, 
\begin{equation}
\lambda^{\rm u} < R_{\rm dec}^{\rm u} = 1.2\times10^{17} \left( \frac{E_{\rm iso, 53}}{n_0^{\rm u}}\right)^{1/3} \Gamma_{\rm sh, 2}^{\rm u\ -2/3}\ {\rm cm}~~,
\end{equation}
where $E_{\rm iso, 53}, n_0^{\rm u}$ and $\Gamma_{\rm sh, 2}^{\rm u}$ are the isotropic equivalent kinetic energy of the blast wave in unit of erg, upstream density in unit of ${\rm cm}^{-3}$ and the initial Lorentz factor of the shock measured in the upstream rest frame, respectively (e.g. \citet{sironi07}). 
We have adopted the usual notation $\mathcal{Q}_n\equiv \mathcal{Q}/10^n$. 
For a later phase of GRB afterglows, the upstream density fluctuations with longer wavelength generate downstream magnetic field.

In the hydrodynamics, a finite-amplitude sound wave eventually steepens to a shock-like structure, and the wave motion is thermalized by dissipation. 
The lifetime of a downstream sound wave with the wavelength of $\lambda_{\rm down}$ in the downstream rest frame is given by \citep{stein72}
\begin{equation}
t_{\rm lifetime} = \frac{\lambda_{\rm down}}{\langle v\rangle}  + t_{\rm dis}~~.
\end{equation}
The first term represents a steepening time in which the sound wave becomes a shock-like structure. 
The second term represents a dissipation time. If the dissipation is mainly due to the shock wave, the dissipation time is given by $t_{\rm dis}=\lambda_{\rm down}/\langle{v\rangle}(M-1)$, where $M$ is a Mach number \citep{stein72}. 
However, at present we have not understood the main dissipation mechanism and cannot estimate the dissipation time because we consider a collisionless system. 
In this discussion, we consider only the steepening time as a lower limit of the lifetime of the downstream sound wave.  
%In order for the upstream density fluctuations to affect the GRB afterglow, the lifetime in the upstream rest frame must be smaller than the deceleration time of the GRB jet, $t_{\rm dec}=R_{\rm dec}/c$, where $t_{\rm dec}/\Gamma^{2}$ corresponds to the peak time of the optical afterglow in observer frame.  
Observed GRB afterglows are already bright until the deceleration time, $t^{\rm u}_{\rm dec}=R^{\rm u}_{\rm dec}/c$, at which  observers see the emission peak in the optical band.
Hence, the field generation occurs until this epoch, so that $\Gamma^{\rm u}_{\rm sh} t_{\rm lifetime}/\sqrt{2}>t^{\rm u}_{\rm dec}$.
Thus, the lower limit of the wavelength of the upstream density fluctuation is given as,
\begin{eqnarray}
 \lambda^{\rm u} > &9.6&\times 10^{16}  \left( \frac{c_s}{0.62c}\right)^{-1} \left( \frac{\langle v \rangle}{0.2c}\right) \nonumber \\
 &\times& \left( \frac{E_{\rm iso, 53}}{n_0^{\rm u}}\right)^{1/3} \Gamma_{\rm sh, 2}^{\rm u\ -2/3}\ {\rm cm}~~, 
\end{eqnarray}
where we use $\lambda^{\rm u}\sim \Gamma_{\rm sh}^{\rm u} \lambda_{\rm down}$ \citep{lemoine16}. 
Since we ignore the dissipation time in this discussion, the above lower limit actually should be smaller.

In our previous local PIC simulation with periodic boundary conditions \citep{tomita16}, 
we investigated effects of particle diffusion from an entropy mode in the shock downstream region. 
We showed that if the spacial structure of the entropy mode is anisotropic, particles anisotropically escape from high-density regions. Then, the temperature anisotropy is generated, so that the magnetic field is generated by the Weibel instability. 
However, the anisotropic escape of particles from the entropy mode was not observed in this shock simulation. 
This is because the strength of the large-scale long-lived magnetic field is large enough to suppress the particle diffusion from the entropy mode. 
Instead of that, high-energy accelerated particles escape from high-density regions of the sound wave, which makes temperature anisotropy in the shock downstream region. 
The large-scale magnetic field generated in the shock precursor region would decay more rapidly in a realistic three-dimensional simulation. 
In that case, the downstream entropy and sound waves would also rapidly dissipate due to the particle diffusion, so that 
the temperature anisotropy would be generated by the anisotropic escape from high-density regions.  
Therefore, a realistic three-dimensional PIC simulation should be performed in the future even though it is very challenging.

In this paper, we study the case where the wave vector of the upstream density fluctuation is parallel to the shock normal.
In reality, there are clumpy structures in the ISM or CSM.
When a high-density clump passes through the shock front, not only compressible modes but also vortexes (incompressible modes) are excited in the downstream region. 
If the particle diffusion works, the escape of particles from the vortex can generate the temperature anisotropy as the sound wave makes it. 
Magnetic-field lines generated by the Weibel instability might be stretched to a large scale by the vorticity. 
Then, the magnetic field would be further amplified by the turbulent dynamo. 
In addition, if the particle diffusion works efficiently, the vortex rapidly decays and the turbulent dynamo becomes inefficient compered with ideal MHD case. 
However, if the shock-compressed background magnetic field is rapidly amplified by the turbulent dynamo before particles escape from the high-density clump, the diffusion is strongly suppressed, so that the small-scale magnetic field cannot be generated by the Weibel instability.  
A larger simulation box and a longer simulation time are required to understand the effect of vortexes on collisionless shocks, which will be addressed in future work.

The decay of turbulence means heating or acceleration of particles by the turbulence. 
Although we cannot see in this study a significant difference of the number of high-energy particles between the homogeneous and inhomogeneous cases, 
the second-order Fermi-acceleration by the downstream large-scale turbulence is expected in larger and longer simulations \citep{ohira13a, asano15}.

 We considered only electron-positron plasma in this study. 
For the standard model of the GRB afterglow, the external shock wave propagates to an electron-ion plasma. 
When the shock is ultra-relativistic, the downstream electron and ion temperatures become almost the same \citep{spitkovsky08a, kumar15}.
Then, the results for electron-positron plasmas would be applicable to GRB afterglows. 
However, the shock eventually becomes non-relativistic at a later phase of the GRB afterglow. 
Therefore, in reality, three-dimensional effects, the upstream magnetic fields, multidimensional density structures, nonrelativistic physics, and ion inertia could play important roles on the GRB afterglow and collisionless shocks propagating to inhomogeneous plasmas. More realistic simulations should be performed to understand these effects. Although it is very challenging for numerical simulations,   we may be able to reveal it in the laboratory astrophysics \citep{takabe08}. 
Furthermore, we can apply their knowledge to supernova remnant shocks which are believed to be the cosmic-ray accelerator in our galaxy. %ラボアスで明らかにできるかも(例えば、高部さんの論文などを引用)。これらの結果は、GRB afterglowだけでなく、Galactic CRの加速器であるSNRのshockにも適応できる。} 

\section{Summary} \label{sec:4}
We have studied unmagnetiezed, relativistic collisionless shocks propagating into inhomogeneous plasmas. 
For the case of inhomogeneous upstream density distribution, large-scale magnetic field is generated by the Weibel instability in the shock precursor region. 
As the large-scale field is advected downstream, 
in contrast to the homogeneous case, they did not decay significantly. 
In this case, the sound and entropy waves can propagate downstream during our simulation time, which produces the downstream temperature anisotropy. 
We found that the upstream inhomogeneity has free energies large enough to explain the field strength inferred from observed GRB afterglows.
%We could expect the effective particle acceleration in the shock propagating into inhomogeneous media.

\acknowledgments
We thank M. Hoshino, Y. Matsumoto, T. Amano for useful comments. 
The software used in this work was in part developed in pCANS at Chiba University. 
Numerical computations were carried out on Cray X50 at Center for Computational Astrophysics, National Astronomical Observatory of Japan. 
This work is supported by Grant-in-Aid for JSPS Research Fellow from Japan Society for the Promotion of Science (JSPS), No. 17J03893 (ST), and in part by JSPS KAKENHI grant numbers 16K17702 (YO), 19H01893 (YO) and 18H01232(RY).

%\appendix
%\section {}\label{appendixA}
%\section {}\label{appendixB}

%% This command is needed to show the entire author+affilation list when
%% the collaboration and author truncation commands are used.  It has to
%% go at the end of the manuscript.
%\allauthors

%% Include this line if you are using the \added, \replaced, \deleted
%% commands to see a summary list of all changes at the end of the article.
%\listofchanges

\end{document}